\documentclass[conference]{IEEEtran}
\ifCLASSINFOpdf
\else
\fi
\usepackage{xcolor}
\usepackage{cite}
\usepackage{amsmath}
\usepackage{amsfonts}
\usepackage[short]{optidef}
\usepackage{subfig}
\usepackage[pdftex]{graphicx}
\usepackage{autobreak}
\newcommand{\T}{\mathcal{T}}
\newcommand{\C}{\mathcal{C}}

\usepackage{todonotes}
\setuptodonotes{inline}
\usepackage{changes}

\begin{document}
%
\title{Planning Mm-Wave Access Networks With Reconfigurable Intelligent Surfaces}

\author{\IEEEauthorblockN{Eugenio Moro$^*$, Ilario Filippini$^*$, Antonio Capone$^*$ and Danilo De Donno$^\dagger$}
\IEEEauthorblockA{
\hfill\\
$^*$ANTLab - Advanced Network Technologies Laboratory, Politecnico di Milano, Milan, Italy \\
$^\dagger$Milan Research Center, Huawei Technologies Italia S.r.l, Milan, Italy\\
Email: $\{$eugenio.moro, ilario.filippini, antonio.capone$\}$@polimi.it, danilo.dedonno@huawei.com
}
}
\maketitle

\begin{abstract}
With the capability to support gigabit data rates, millimetre-wave (mm-Wave) communication is unanimously considered a key technology of future cellular networks. However, the harsh propagation at such high frequencies makes these networks quite susceptible to failures due to obstacle blockages. Recently introduced Reconfigurable Intelligent Surfaces (RISs) can enhance the coverage of mm-Wave communications by improving the received signal power and offering an alternative radio path when the direct link is interrupted. While several works have addressed this possibility from a communication standpoint, none of these has yet investigated the impact of RISs on large-scale mm-Wave networks. Aiming to fill this literature gap, we propose a new mathematical formulation of the coverage planning problem that includes RISs. Using well-established planning methods, we have developed a new optimization model where RISs can be installed alongside base stations to assist the communications, creating what we have defined as Smart Radio Connections. Our simulation campaigns show that RISs effectively increase both throughput and coverage of access networks, while further numerical results highlight additional benefits that the simplified scenarios analyzed by previous works could not reveal.
\end{abstract}

\IEEEpeerreviewmaketitle

\section{Introduction}
\label{sec:intro}
Current and future mobile radio network generations are challenged to cope with ever-expanding mobile data demands, spurred by our increasingly connected society~\cite{rapparport2014}.\newline 
At the same time, cellular communication systems based on sub-6GHz frequencies are currently experiencing a bandwidth shortage~\cite{rappaport2019} as they struggle to deliver the required level of performance.\newline 
Millimetre-wave (mm-wave) based cellular communications have been recognized as the key technology to address both these crucial issues, as they can fulfil the promise of supporting Gbps demands while also solving the spectrum scarcity issue~\cite{rapparport2013}.\newline
Although its standardization in cellular networks for mobile access began only recently with 3GPP Release 15, this technology has already been largely employed in satellite links and cellular backhauling~\cite{ericssonBackhaul} and its limitations are well known.  
In particular, mm-waves are affected by harsh propagation typical of such high frequency that leads to high free space attenuation. Simultaneously, high penetration losses and poor diffraction mean that any obstacle crossing the line of sight might easily cause mm-Wave communications to fail.\newline
While emergent technologies - such as massive MIMO and beamforming - can effectively compensate for the increased pathloss~\cite{kutty2016}, the problem of blockage resiliency in mobile access has not encountered the same luck.

Among the candidate technologies that can potentially address the issue above, the recent emerging concept of Reconfigurable Intelligent Surface (RIS) has gained extreme popularity among the academic community~\cite{basar2019}.\newline
RISs are described as quasi-passive planar structures whose electromagnetic properties can be electronically controlled to manipulate impinging radio waves in a variety of ways. While an RIS can produce several types of these electromagnetic manipulations, the ability to reflect and focus impinging waves in any direction has the potential of transforming these surfaces in passive relays~\cite{DiRenzo2020}. This ability is exciting for mm-Wave communications, as an RIS can increase the blockage resilience by creating an alternative \textit{electromagnetic path}. As opposed to active relays, RISs also show significantly higher energy efficiency~\cite{huang2019reconfigurable} and prototypal works~\cite{tan2018} have shown how they can be effectively built with cheap materials. Indeed, part of the attention that RISs are generating might be well justified by the opportunity of reducing the cost of deploying and maintaining a resilient wireless access network as opposed to more traditional and expensive approaches~\cite{devoti2020}.

Theoretical works~\cite{Cao2020}\cite{Wang2019}\cite{du2021} have extensively analyzed this particular RIS configuration from a communication perspective, providing practical mathematical tools to model the propagation characteristics of such scenarios. However, these analyses are carried out at the link level with simplified network scenarios.\newline
In this work, instead, we focus on the planning of large-scale mm-Wave radio access networks employing intelligent surfaces and, to the best of our knowledge, it is the first to tackle this challenge.  We have employed well-established coverage planning methods to develop a new mathematical formulation of the coverage planning problem where both base stations and RISs can be installed in given locations of an arbitrary geographic area. We have introduced the concept of~\textit{Smart Radio Connection} (SRC), a logical abstraction of the well-known concept of the RIS-enabled \textit{Smart Radio Environment}~\cite{di2019smart}. An SRC consists of a radio link assisted by an intelligent surface and, in our planning model, SRCs can be established alongside traditional connections between UEs and base stations to increase the coverage and system performance.\newline
Our extensive numerical analysis campaign testifies how the well known point-to-point benefits of employing RISs do scale well at the system level for mobile access. Results show that including RISs when planning a radio access network can simultaneously increase coverage, throughput and blockage resiliency. Additionally, our results give new interesting insights on the benefits of employing RISs for coverage planning of mm-wave networks that could not be noticed in the highly simplified scenarios of related works. In particular, our model can identify the RIS configurations and the deployment budget conditions that provide tangible performance advantages when RISs are considered.

The rest of this paper is structured as follows: Sec.~\ref{sec:related_works} presents some relevant related works, Sec.~\ref{sec:base_model} details a baseline mm-wave coverage planning model that does not include the presence of RISs, Sec.~\ref{sec:ris_model} describes the modeling choices that lead us to develop a RIS-aware planning model and presents the novel mathematical formulation. Finally, Sec.~\ref{sec:results} shows the simulation setup and the numerical results.

%
\section{Related Works}
\label{sec:related_works}
Reconfigurable Intelligent Surfaces represent the latest technological proposition in the domain of propagation waves control~\cite{alexandropoulos2020reconfigurable}. Their use as passive relays has been proposed in~\cite{DiRenzo2020}, where preliminary link-level simulations have shown the potential benefits with respect to more traditional active relaying approaches.\newline
From a communication standpoint, the problem of jointly optimizing the base station pre-coding and the RIS elements phase shifts has been studied in~\cite{Cao2020}, where an iterative algorithm addresses the non-convexity challenges. In~\cite{Wang2019}, a closed-form solution of the same problem is derived exploiting the characteristic of mm-Wave channels.\newline
Finally, authors of~\cite{tan2018} have shown how a prototype RIS can effectively enhance the coverage of indoor mm-Wave networks.

Historically, the problem of coverage planning has been applied to different radio access technologies.\newline
However, mm-Wave coverage planning works have only lately appeared in the literature, given the relatively recent interest. Understandably, these works have studied the coverage problem with a focus on the network resilience against blockages.\newline
In particular, authors of~\cite{palizban2017} study the problem of optimizing the layout of an mm-Wave access network in a dense urban environment such that the LOS availability is maximized. A similar analysis is carried out in~\cite{mavromatis2019efficient} for mm-Wave vehicular communication scenarios.\newline 
In~\cite{wu2017}, the coverage planning problem is studied through a network cost minimization that employs a link availability stochastic model. Finally, authors of~\cite{devoti2020} have studied the impact of different network planning approaches on the blockage resiliency of mm-Wave deployments.\newline
None of the planning works mentioned above has included reconfigurable intelligent surfaces in their investigations. To the best of the authors' knowledge, this is the first published work to present such an analysis. 
\section{Basic mm-Wave Model}
\label{sec:base_model}
In this section, we give a basic description of a mathematical programming model for mm-Wave access network coverage planning. Similarly to other coverage planning works~\cite{AMALDI20082159,devoti2020}, we identify a set $\C$ of candidate positions (i.e. Candidate Sites, CSs) over a given geographic area where Base Stations (BS) can be installed. A discrete set of Test Points (TP) $\T$ represents the traffic/user distribution.\newline
Binary coverage parameter $\Lambda_{t,c}$ captures the propagation characteristics between TP $t \in \T$ and CS $c \in \C$. Particularly, $\Lambda_{t,c}=1$ if a radio link between the two positions can be established and zero otherwise. These parameters are set according to physical considerations, such as distance, transmission power, receiver sensitivity, antenna gain, attenuation losses, and more. Additionally, blockages due to fixed and opaque obstructions between any pair of CS-TP can be modelled by setting the corresponding coverage parameter to 0.\newline
Given the fixed known position of any potential CS-TP pair, the maximum achievable downlink bit-rate can be pre-computed according to the transmitter and receiver characteristics and any propagation model of choice. Indeed, given the extreme directivity of mm-Wave downlink transmissions that can strongly limit any interference effect, we can reasonably assume this bit-rate to be independent of other simultaneous access transmissions~\cite{devoti2020}.\newline
However, a well-known issue of millimetre-based communication is its high penetration loss and limited diffraction~\cite{gerasimenko2019}, resulting in frequent blockages due to obstacles transiting across the connection line of sight. Blocked radio links experience a dramatic reduction in throughput, and this can be taken into consideration by weighting the maximum achievable bit-rate of each link with the probability of the link being in a state where such bit-rate is actually available (i.e., not blocked)\footnote{Specific blockage models, such as~\cite{Akdeniz2014}, express this probability as a decreasing function of the link length, allowing this quantity to be computed given the CS-TP distances.}.
Parameter $R_{t,c}^\text{BS}$ denote this expected (\textit{blockage-weighted}) maximum throughput between TP $t \in \T$ and BS installed in $c \in \C$. Similarly, $R^\text{MIN}$ identifies the minimum expected throughput that needs to be guaranteed to each TP for it to be considered as covered. Knowing the channel states $\mathcal{S}$, their probabilities $p_{s},s \in \mathcal{S}$ and the corresponding achievable rates $r_{s}, s \in \mathcal{S}$, these parameters can be computed according to the following formula:
\begin{equation}
    R = \sum_{s \in \mathcal{S}}p_{s}r_{s}.
\end{equation}
Finally, the coverage planning is constrained to a budget value $B$ and parameter $P_c$ describes the cost of installing a BS in a particular CS $c \in \C$.\newline
The proposed planning model is based on the following decision variables:
\begin{itemize}
    \item $y_c^\text{BS} \in \{0,1\}$: installation variable equal to 1 if a BS is installed in site $c \in C$ and 0 otherwise,
    \item $x_{t,c} \in \{0,1\}$: association variable equal to 1 if BS in $c \in \C$ is assigned for coverage of test point $t \in \T$,
    \item $\tau_{t,c}^\text{BS} \in [0,1]$, time-sharing variable indicating the fraction of time during which BS in $c \in \C$ transmits to test point $t \in \T$. This variable allows us to model the BS resource sharing as a time-sharing process, in accordance to 3GPP Rel. 15 specifications. Note that the very same notation can be applied if the joint time and sub-carrier sharing has to be considered.
\end{itemize}
Given the notation, the parameters and the variables described above, we now propose a basic MILP (Mixed Integer Linear Programming) formulation of the coverage planning problem:
\begin{maxi!}[2]
  {}{\sum_{t \in \T, c \in \C} R_{t,c}^\text{BS}\cdot\tau^\text{BS}_{t,c}}{}{}\label{opt1:obj}
  \addConstraint{\sum_{c \in \C}x_{t,c}}{\leq 1}{\forall t \in \T}\label{opt1:1donor}
  \addConstraint{\tau_{t,c}^\text{BS}}{\leq \Lambda_{t,c}\cdot x_{t,c}\quad}{\forall t \in \T, c \in \C}\label{opt1:tau_act}
  \addConstraint{\sum_{t \in \T}\tau_{t,c}^\text{BS}}{\leq y_c^\text{BS}}{\forall c \in \C} \label{opt1:tdm}
  \addConstraint{\sum_{c \in \C} R_{t,c}^\text{BS}\cdot\tau_{t,c}^\text{BS}}{\geq R^\text{MIN}\quad}{\forall t \in T}\label{opt1:min_rate}
  \addConstraint{\sum_{c \in \C}{P_c\cdot y_c^\text{BS}}}{\leq B}\label{opt1:budget}
\end{maxi!}
The objective function in~(\ref{opt1:obj}) expresses the goal of the planning model: the maximization of the sum-throughput. A per-user average throughput appears in the sum, which depends on both the nominal link capacity between BS and TP and the fraction of resources the BS dedicates to the specific TP. Also, note that we consider this objective function as one of the very many possible ones. Other approaches, such as the sum of throughput logarithms, the max-min throughput, etc., can be easily plugged in with minimal changes to the formulation.\newline
Constraints~(\ref{opt1:1donor}) enforces each TP to be covered at most by 1 BS. Constraint~(\ref{opt1:tau_act}) is such that a BS in $c \in \C$ can transmit to a TP $t \in \T$ for a strictly positive fraction of time only if such TP is associated with this particular BS (i.e. $x_{t,c} = 1$) and if a radio link can be established between the two (i.e. $\Lambda_{t,c} = 1$).\newline
Constraint~(\ref{opt1:tdm}) has a double function. First, it does not allow any transmission of strictly positive duration to originate from any BS which has not been installed. Additionally, it limits to 1 the overall sum of the fractions of time dedicated for transmissions towards specific TPs for each installed BS, effectively enforcing a time-based division of BS resources. Note that this constraint may imply single-beam BS transmissions. However, the goal of this formulation is not to provide a perfect user throughput figure, which is usually computed by system-level simulators, but rather to design a good network layout. The latter can be achieved even with approximated user throughput models that do not substantially change the optimal deployment. On top of that, multi-beam antenna patterns remarkably decrease link directivity, strongly limiting BS coverage. As such, we believe it is reasonable to assume that most of the downlink transmissions involve one user at a time.\newline
Constraint~(\ref{opt1:min_rate}) simply bounds each TP's throughput to be at least the minimum throughput $R^\text{MIN}$.\newline
Finally, constraint~(\ref{opt1:budget}) limits the deployment cost to the available planning budget $B$, with $P_c^\text{BS}$ indicating the cost of installing a BS in CS $c \in \C$.
%
\section{Modelling Reconfigurable Intelligent Surfaces}
\label{sec:ris_model}
\begin{figure}[t]
    \centering
    \includegraphics[width=0.5\columnwidth]{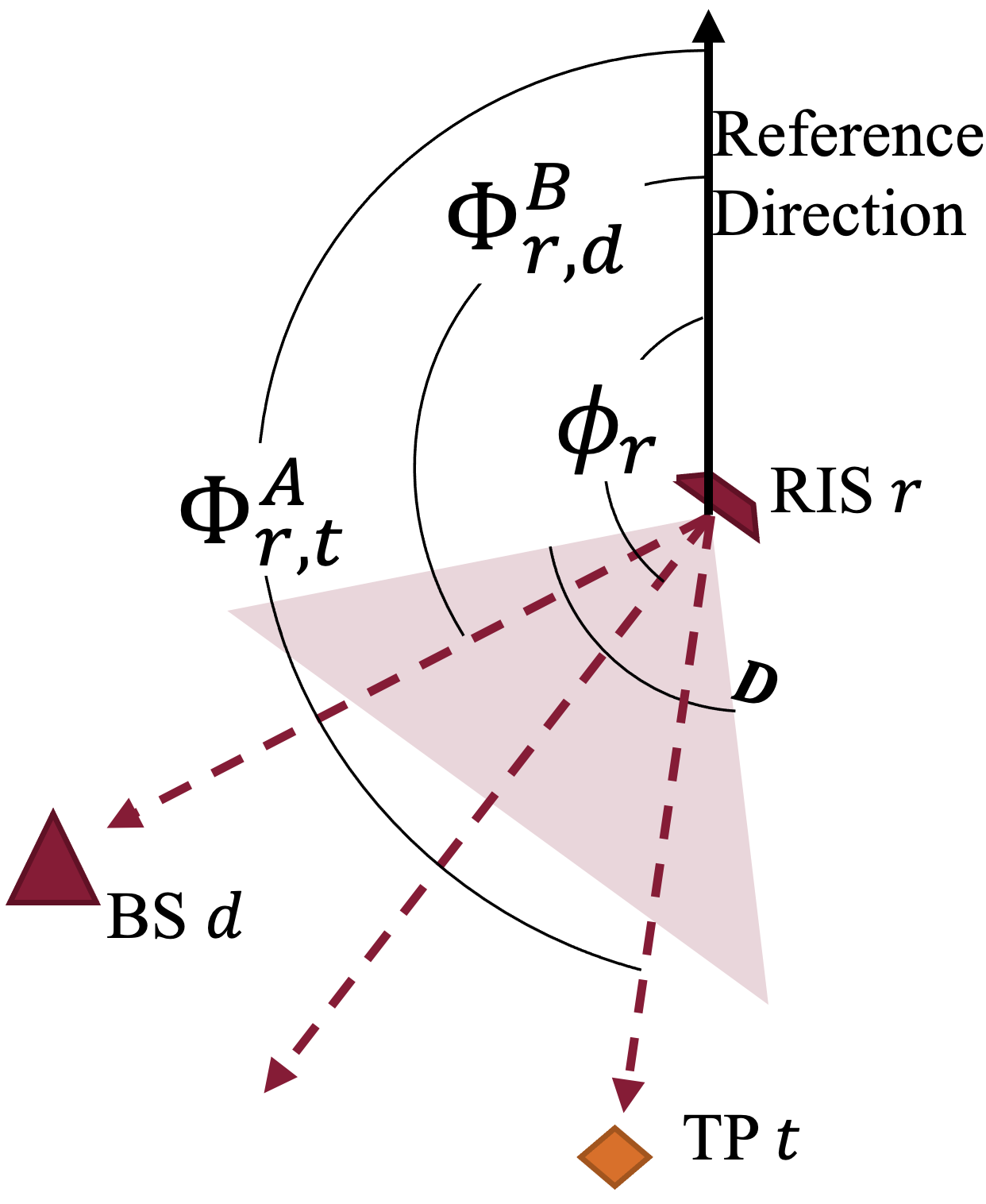}
\caption{Example of SRC with RIS orientation and lines of sight angles.}
\label{fig:src}
\end{figure}In our modelling efforts, RISs behave as \textit{passive beamformers}, focusing the impinging radio waves in specific directions and creating what is often identified as a \textit{Smart Radio Environment}. In this way, a proper configuration of the RIS can actively assist the communication between a transmitter-receiver pair by increasing the Signal to Noise Ratio (SNR) at the receiver~\cite{DiRenzo2020}. Following the same rationale, we introduce the novel concept of \textit{Smart Radio Connection} (SRC): a triplet that comprises one transmitter (i.e. the BS), one receiver (i.e. the UE located in a specific TP) and a smart surface configured to assist this specific communication\footnote{While it is possible for  multiple RIS to be configured to assist a single TX-RX pair~\cite{Cao2020}, in this work we focus on up to one surface per SRC.}. Any SRC is then modeled as a tuple $<t,d,r>$, where $t \in \T$ denotes the TP, $d \in \C$ denotes the BS installation site and $r \in \C$ denotes the RIS installation site, as the example pictured in Figure~\ref{fig:src} shows.\newline
The problem of jointly optimizing the transmitter pre-coding and the RIS elements' phase shifts in a SRC is generally not convex~\cite{Cao2020}. However, the inherent characteristics of a mm-Wave channel allow for significant simplifications and an optimal closed form expression of the average received power can be derived. In this work, we consider the average SRC channel gain expression developed in~\cite{Wang2019} for mm-Wave communication, which we propose here in a compact form:
\begin{equation}
    \label{eq:channel}
    \gamma = f(\mathbf{h}_{B,R},\mathbf{h}_{R,P}) + f'(\mathbf{h}_{B,R},\mathbf{h}_{R,P}, \mathbf{h}_{B,P}) + f''( \mathbf{h}_{B,P}),
\end{equation}
where $\mathbf{h}_{B,R}$ is the channel between the BS and the RIS, $\mathbf{h}_{R,P}$ is the channel between the RIS and the TP, $\mathbf{h}_{B,P}$ is the channel between the BS and the TP and $f,f',f''$ are proper functions.\newline
The contribution of the RIS to the SRC channel gain is linearly separable from the contribution of the traditional direct link, meaning that the increment in SRC link capacity with respect to unassisted communication is directly dependent only on the terms $f(\mathbf{h}_{B,R},\mathbf{h}_{R,P})+f'(\mathbf{h}_{B,R},\mathbf{h}_{R,P}, \mathbf{h}_{B,P})$. It follows that, by knowing the relative positions of the three components of a SRC, as well as the state probability of each channel, the performance of any SRC can be completely characterized. Indeed, we define $R_{t,d,r}^\text{SRC}$ as the expected (\textit{blockage-weighted}) throughput when BS in $d \in \C$ transmits to TP $t \in \T$, while being assisted by RIS in $r \in \C$.\newline
In general, a RIS can be part of many SRCs, and we assume an instantaneous reconfiguration of the reflecting elements when the surface switches between different SRCs. However, we allow each surface to assist up to 1 TX-RX pair at a time, meaning that the RIS sharing takes the form of a time-sharing process.\newline 
We are fully aware that the previous assumptions may represent some though technological challenges for RIS hardware manufacturers. However, we believe them to be consistent with a realistic technological maturity level that needs to be considered from the beginning if we want to investigate the potential benefits of RIS development. For instance, a similar evolution occurred in literature to beamforming reconfiguration assumptions.\newline 
Similarly to what happens for uniform linear antenna arrays, RISs are expected to present a limited array field of view~\cite{tan2018}. We consider this by defining a RIS orientation, coinciding with the vector normal to the surface. For a given orientation, the lines of sight of the base stations/test points of all SRCs which the RIS is assigned to have to fall inside the surface field of view. In this work, we define a horizontal field of view angle $D$ and we discard the vertical field of view\footnote{It usually has a limited impact on the network layout, however, if needed, a vertical field of view can be easily included in the model}.\newline
Finally, our proposed model maintains generality by not forcing any BS-TP pair to be RIS-assisted. However, including both SRCs and traditional direct-link radio connections in a planning model was found to require a cumbersome number of additional variables and constraints. We worked around this issue by including an additional candidate site $\Tilde{c}$ where a \textit{fake} RIS is always installed. This particular RIS has no cost, no time-sharing limitation and 360° field of view, but grants no additional throughput performance to any assisted BS-TP pair. After an optimal solution is found, a post-processing operation changes any SRC including the \textit{fake} RIS into a traditional unassisted BS-TP communication. This way, we could maintain a leaner formulation by modelling SRCs only, while avoiding any loss of generality.\newline
According to the previously described modeling choices, the following variables were needed to extend the mm-Wave coverage planning model presented in sec.~\ref{sec:base_model}:
\begin{itemize}
    \item $y_c^\text{RIS}\in \{0,1\}$: RIS installation variable, equal to 1 if a RIS is installed in site $c\in \C$ and 0 otherwise,
    \item $s_{t,d,r}\in \{0,1\}:$ SRC activation variable, equal to 1 if RIS in $r \in \C$ is assigned to assist the communication between BS in $d \in \C$ and TP $t \in \T$,
    \item $\tau_{t,d,r}^\text{SRC} \in [0,1]:$ SRC time sharing variable, indicating the fraction of time during which BS in $d \in \C$ transmits to TP $t \in \T$ aided by a RIS installed in $r \in \C$,
    \item $\phi_r \in [0, 2\pi]:$ azimuth of RIS installed in CS $r \in \C$ computed with respect to a reference direction.
\end{itemize}
We are now ready to introduce the coverage planning model extended to include Reconfigurable Intelligent Surfaces:
\begin{maxi!}[3]
  {}{\sum_{t \in \T, d \in \C, r \in \C} R_{t,d,r}^\text{SRC}\cdot\tau^\text{SRC}_{t,d,r}}{}{}\label{opt2:obj}
  \addConstraint{y_c^\text{BS}+y_c^\text{RIS}}{\leq 1}{\forall c \in \C}\label{opt2:install}
  \addConstraint{y_{\Tilde{c}}^\text{RIS}}{\geq 1} \label{opt2:fakeris}
  \addConstraint{\sum_{d \in \C, r \in \C}s_{t,d,r}}{\leq 1}{\forall t \in \T}\label{opt2:1src}
  \addConstraint{\tau_{t,d,r}^\text{SRC}}{\leq \Lambda_{t,d,r}\cdot s_{t,d,r}}{\forall t \in \T,\, d,r \in \C}\label{opt2:tau_act}
  \addConstraint{\sum_{t \in \T, r \in \C}\tau_{t,d,r}^\text{SRC}}{\leq y_d^\text{BS}}{\forall d \in \C} \label{opt2:bs_tdm}
  \addConstraint{\sum_{t \in \T, d \in \C}\tau_{t,d,r}^\text{SRC}}{\leq y_r^\text{RIS}}{\forall r \in \C \setminus\{\Tilde{r}\}} \label{opt2:ris_tdm}
  \addConstraint{\sum_{d \in \C, r \in \C}R_{t,d,r}^\text{SRC}\cdot\tau^\text{SRC}_{t,d,r}}{\geq R^\text{MIN}}{\forall t \in T}\label{opt2:min_rate}
  \addConstraint{\phi_r}{\geq \Phi^{\text{A}}_{r,t} - D/2 - 2\pi(\neg s_{t,d,r})}{\forall t \in \T, d,r \in \C: r \neq \Tilde{c}}\label{opti2:or1}
  \addConstraint{\phi_r}{\leq \Phi^{\text{A}}_{r,t} + D/2 + 2\pi(\neg s_{t,d,r})}{\forall t \in \T, d,r \in \C: r \neq \Tilde{c}}\label{opti2:or2}
  \addConstraint{\phi_r}{\geq \Phi^{\text{B}}_{r,d} - D/2 - 2\pi(\neg s_{t,d,r})}{\forall t \in \T, d,r \in \C: r \neq \Tilde{c}}\label{opti2:or3}
   \addConstraint{\phi_r}{\leq \Phi^{\text{B}}_{r,d} + D/2 + 2\pi(\neg s_{t,d,r})}{\forall t \in \T, d,r \in \C: r \neq \Tilde{c}}\label{opti2:or4}
  \addConstraint{\sum_{c \in \C\setminus \{\Tilde{c}\}}{\left(P_c^\text{BS}\cdot y_c^\text{BS} + P_c^\text{RIS}\cdot y_c^\text{RIS}\right)}}{\leq B\hspace{-5mm}}\label{opt2:budget}
\end{maxi!}
Objective function~(\ref{opt2:obj}) is of the sum-throughput type. Constraint~(\ref{opt2:install}) makes sure that a BS and a RIS cannot be installed in the same candidate site, while~(\ref{opt2:fakeris}) forces the installation of the \textit{fake} surface. Constraint~(\ref{opt2:1src}) allows for up to 1 SRC to be active for each TP, meaning that each $t \in \T$ is covered by up to 1 BS and up to 1 RIS. In~(\ref{opt2:tau_act}-\ref{opt2:ris_tdm}) the BS and RIS time sharing is enforced. In particular, a strictly positive transmission duration is allowed only if the SRC is active, if both BS and RIS are installed and if a radio connection between the three network components can be established\footnote{Note that, while the coverage parameter $\Lambda_{t,d}$ has been extended to also include a third index representing the RIS CS, its rationale remains unchanged.}. Constraints~(\ref{opti2:or1}-\ref{opti2:or4}) force the RIS azimuth to be such that the lines of sight of any associated BS and TP all fall inside its field of view. Parameters $\Phi_{r,t}^\text{A}$ and $\Phi_{r,d}^\text{B}$ indicate the angle between a reference vector originating from RIS $r \in \C$ and the connected TP $t\in \T$ and BS $d \in \C$ lines of sight, respectively. The reader can find an illustration in Figure~\ref{fig:src}. Note that $\neg s_{t,d,r}= (1-s_{t,d,r})$. Finally, we have introduced a RIS cost parameter $P_c^\text{RIS}$ in the budget constraint~(\ref{opt2:budget}).


%
\section{Results}
\label{sec:results}
In this section, we numerically analyze the previously described models when applied to different instances. Such instances are characterized by parameters that vary according to the specific result or property intended to be highlighted. However, some assumptions will be valid throughout the entire section unless otherwise stated.\newline
We consider scenarios where the BS employs several uniform linear antenna arrays, such that the BS field of view is 360°. We assume 64 antennas per array and a transmit power of $30dBm$. The receiver's antenna is assumed to be omnidirectional, and RX sensitivity is set to $-78dBm$.\newline
Given that the size of the reflecting surface is directly related to the system performance~\cite{Bjornson2020}, we show results for both $10^4$ and $10^5$ reflecting elements in each RIS. These are compatible with surface sizes of about $50\text{x}50cm$ (i.e. small RIS) and $150\text{x}150cm$ (i.e. large RIS), respectively, since the reflecting elements need around $\lambda/2$ spacing~\cite{Renzo2020}. Additionally, RIS field of view is set to 120°.\newline
Carrier frequency is set to $28GHz$ and both propagation and blockage models are taken from~\cite{Akdeniz2014}. According to this model, the expected throughput decreases with the link-length, as longer links incur in higher blockage probabilities.\newline 
The received power of SRCs has been computed with the formula derived in~\cite{Wang2019} and summarized by Eq.~\ref{eq:channel}. Traditional direct communication received powers have been computed using the same formula, but discarding the RIS contributions, without loss of generality.\newline
Maximum achievable bit-rates are computed according to realistic modulation and coding schemes, like those specified by IEEE 802.11ad standard~\cite{sur2015}.\newline
In each instance, 52 CSs and 32 TPs are randomly but uniformly scattered on a $400\text{x}300m$ area.\newline
The default planning budget is set to $10.6$. BS cost is set to 1, while large and small RIS costs are set to $0.1$ and $0.05$, respectively.\newline
For any given set of parameters, numeric results have been computed by averaging on 30 random instances of TP and CS positions.\newline
We have used MATLAB to generate each instance and CPLEX to find an optimal solution.

\begin{figure}[t]
    \centering
    \includegraphics[width=0.75\columnwidth]{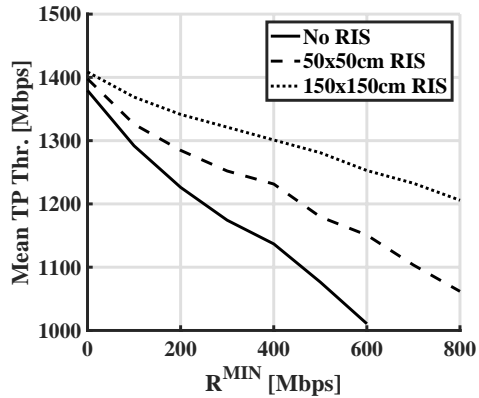}
\caption{Mean TP throughput varying $R^\text{MIN}$}
\label{fig:tp_vs_rmin}
\end{figure}
The first result we intend to analyze is the performance in terms of expected throughput experienced at the test points for different values of $R^\text{MIN}$.\newline 
Figure~\ref{fig:tp_vs_rmin} shows this value averaged over all TPs, for $R^\text{MIN}$ spanning from $0Mbps$ to $800Mbps$, with $100Mbps$ increments.\newline
We note how, independently on the RIS size, the basic planning model is outperformed by the model that includes intelligent surfaces for any value of $R^\text{MIN}$.\newline 
Additionally, larger surfaces perform better than their smaller versions, and the performance difference between the 3 cases grows with the minimum guaranteed throughput. This suggests that the well studied link-level benefits of employing RIS in mm-Wave communication scale well also at system-level.\newline
Finally, the model without RIS becomes unfeasible when $R^\text{MIN} > 600 Mbps$, while optimal solutions can still be found for both RIS sizes. This shows how re-configurable surfaces allow mm-Wave radio access networks to go beyond the coverage capabilities of traditional networks when a larger minimum guaranteed throughput is required.

\begin{figure*}[!t]
    \centering
    \subfloat[Mean TP throughput]{\includegraphics[width=2.6in]{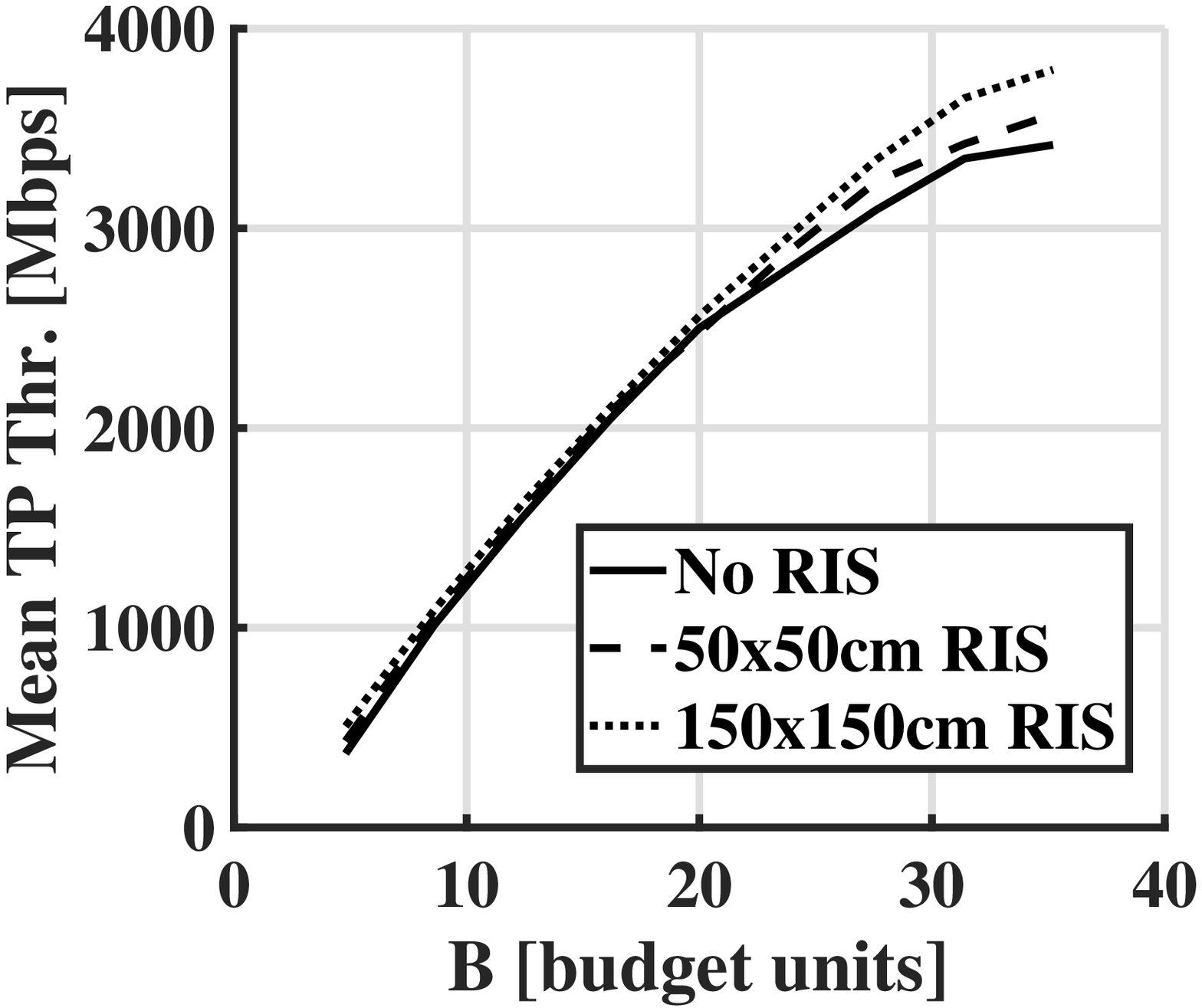}%
\label{fig:rate}}
\hfil
\subfloat[Active sites ]{\includegraphics[width=4.0in]{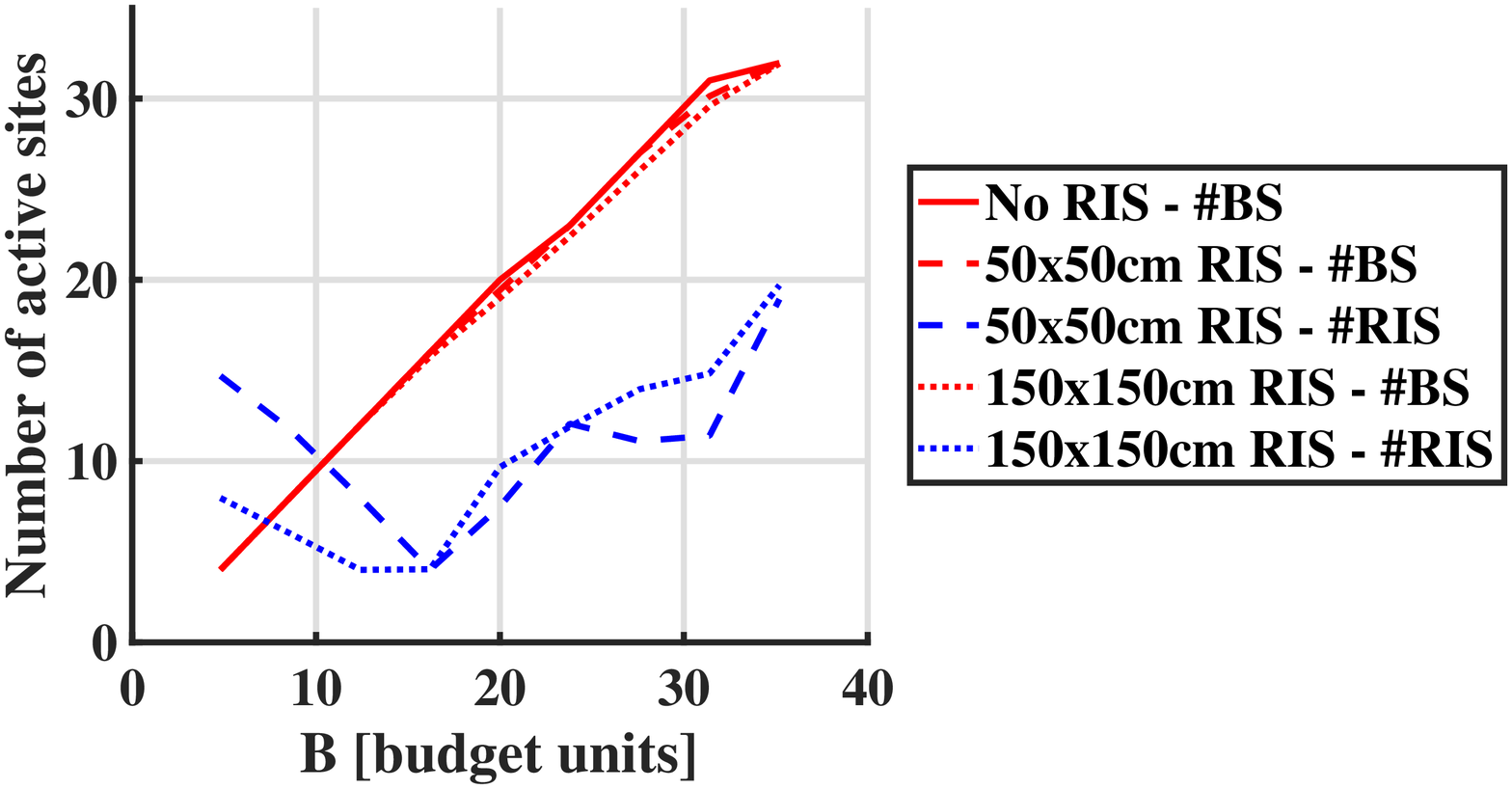}%
\label{fig:active_sites}}
\caption{Budget variations, from $5$ to $35$ units with $4$ units increments.}
\label{fig:tp_vs_b}
\end{figure*}
\begin{figure}[t]
    \centering
    \includegraphics[width=0.8\columnwidth]{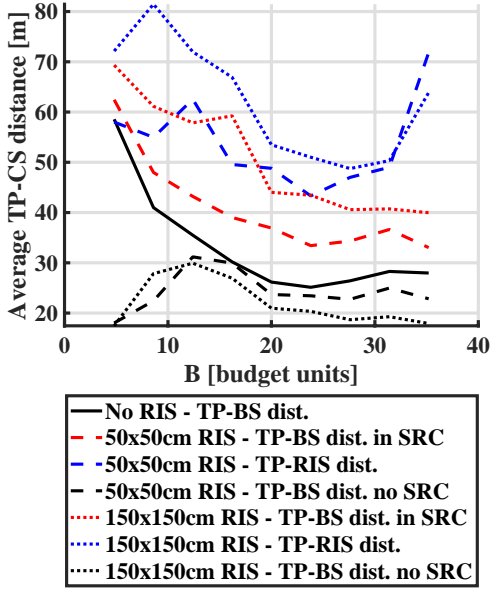}
\caption{Average TP-CS distances when varying budget.}
\label{fig:dist_vs_b}
\end{figure}
We have shown how intelligent surfaces effectively augment the coverage while also increasing the TP experienced throughput. In the following results, we further expand the analysis of the latter in order to establish the efficacy of RISs in boosting the raw network performance.\newline
We set $R^\text{MIN}=100Mbps$ and let $B$ span from around $6$ units to around $36$ units with increments of $4$ units. Note that $B=36$ is equivalent to an infinite budget since it allows the installation of the maximum number of BSs and RISs given the other parameters.\newline
Figure~\ref{fig:rate} shows the impact of the available budget on the experienced TP throughput, while in Figure~\ref{fig:active_sites} we have plotted the variations in the number of active sites where either BSs or RISs are installed.\newline
Interestingly, the number of active sites where RISs are installed - dotted and dashed blue curves in Figure~\ref{fig:active_sites} - decreases as the budget increases from $4$ until around $16$ units, independently on the RIS size. For the same values of $B$, the number of installed base stations increases.\newline
Optimal solutions for lower budgets seem to favour a relatively larger number of RIS installations, which is reduced when BSs substitute RISs as more budget becomes available. However, while still being able to provide adequate coverage levels, the larger count of RISs has little impact on performance boosting for such low values of $B$, as Figure~\ref{fig:rate} testifies.\newline
Indeed, this figure shows that a budget of 20 units or more is needed in order to experience a more substantial raw performance boost, which also coincides with an increased installed RISs count. The suggestion is that the sites where to install intelligent surfaces are chosen to increase coverage for lower values of $B$, while, as the budget increases, additional RISs are installed to increase the throughput.\newline
We confirm this by showing the average TP-RIS distances - dashed and dotted blue curves - against the budget variations in Figure~\ref{fig:dist_vs_b}. Here is indeed evident how these distances decrease at first, as the budget increases, testifying that RISs are installed closer and closer to TPs in order to decrease the probability of blockage and thus guarantee a better coverage.\newline
However, when $B\geq 32$ units, the average distances abruptly increase together with the average RIS installation count. Note indeed that only up to $32$ BSs can be installed (i.e. one per TP), leaving the remaining budget to be spent entirely on RIS installations.\newline
This behaviour indirectly shows how RISs are most effective in boosting the radio access network performance when a portion of the planning budget can be dedicated to their installation or, in other words, when the BSs have been already installed. This is arguably an exciting result, as it suggests that intelligent surfaces might be quite effective in boosting the performance of mm-Wave access networks that have been already deployed.

We conclude this section by providing additional comments on Figure~\ref{fig:dist_vs_b}.\newline
Consider the solid black line and both the dashed and dotted red lines. These represent the average optimal TP-BS distance for the model without RISs (solid black), the average optimal TP-BS distance in SRCs with small RIS size (dashed red) and the same quantity for larger RIS size (dotted red).\newline
In general, we can expect SRCs to be more robust against blockages because multiple lines of sight need to be interrupted at the same time for the connection to fail.\newline 
This concept becomes evident when comparing the 3 curves above, as they show how base stations belonging to SRCs can be placed further away from the test points without reducing the \textit{blockage-weighted} throughput as opposed to BS-TP distances found by solving the base model. Additionally, SRCs allow for a more efficient BS resource sharing, since on average more TPs are in the coverage range of each BS.\newline
As mentioned in Section~\ref{sec:ris_model}, the RIS-aware model still allows for any TP to be covered by a traditional connection if such a choice is optimal. In this regard, the dashed and dotted black curves in Figure~\ref{fig:dist_vs_b} show how those TPs which are covered through a traditional connection are, on average, remarkably closer to the assigned BS with respect to the test points involved in SRCs. This confirms that optimal TP-RIS assignments are chosen such that TPs which are further away from base stations are prioritized, while also suggesting that a heuristic approach based on such policy might yield satisfying results. 
\section{Conclusion}
To study the effect of RISs on large-scale mm-Wave access networks, we have developed a new mathematical formulation for the coverage planning problem that includes reconfigurable surfaces. In our models, RIS can be installed in given candidate sites of a geographic area to assist the communication between base stations and test points, effectively creating what we call a \textit{Smart Radio Connection}. We have also formulated a baseline model where the coverage planning does not consider the presence of RISs. Our simulation campaigns show how RISs can effectively increase both performance and throughput of access networks. Numerical results also highlight the impact of the planning budget on the KPIs above. In particular, we have shown how RISs can offer better coverage even for relatively low budget values, while increasingly noticeable throughput gains are obtained for larger values. Finally, our analysis on the optimal distances between base stations, RISs and test points have shown which RIS positioning policies are the most effective. The study of different planning objectives, more complex deployment scenarios and more refined channel models might be subject of future works. 
\section*{Acknowledgment}
The research in this paper has been carried out in the framework of
Huawei-Politecnico di Milano Joint Research Lab. The Authors
acknowledge Huawei Milan research center for the
collaboration.



%



\bibliographystyle{IEEEtran}
\bibliography{biblio.bib}

\end{document}